\pdfoutput=1
\documentclass[11pt,a4paper]{article}

\usepackage[T1]{fontenc}
\usepackage[utf8]{inputenc}
\usepackage{amsmath}
\usepackage{XCharter}
\usepackage[charter,vvarbb,scaled=1.05]{newtxmath}
\IfFileExists{sourcesans.sty}{\usepackage[scale=0.93]{sourcesans}}{\usepackage[scale=0.93]{sourcesanspro}}
\usepackage[a4paper,top=2.4cm,bottom=2.6cm,left=2.5cm,right=2.5cm,headsep=0.8cm]{geometry}
\usepackage{microtype}
\usepackage{graphicx}
\usepackage{booktabs}
\usepackage{array}
\usepackage{siunitx}
\usepackage[labelfont={bf,sf},font=small,labelsep=period]{caption}
\usepackage{subcaption}
\usepackage{xcolor}
\usepackage{algorithm}
\usepackage{algpseudocode}
\usepackage{enumitem}
\usepackage{titlesec}
\usepackage{fancyhdr}
\usepackage[numbers,sort&compress]{natbib}
\definecolor{linkblue}{HTML}{1F5FAD}
\usepackage[pdfusetitle,colorlinks=true,linkcolor=linkblue,citecolor=linkblue,urlcolor=linkblue,breaklinks=true]{hyperref}
\usepackage[nameinlink,capitalise]{cleveref}
\crefname{equation}{Eq.}{Eqs.}
\crefname{algorithm}{Algorithm}{Algorithms}

\hypersetup{
  pdftitle={Machine learning interatomic potentials and orientational defects in polymers},
  pdfauthor={A. Kiv, V. Soloviev, A. Tuzhykov, A. Bielinskyi, T. Kavetskyy, N. Hoivanovych, O. Sausa, J. Rusnak, H. Svajdlenkova, J. Ostrauskaite, M. Slusarenko},
  pdfkeywords={acrylated epoxidized soybean oil, photopolymerization, dielectric relaxation, Shannon entropy, correlation integral, orientational defects, activation energy, machine learning interatomic potentials, MACE}
}

\titleformat{\section}{\sffamily\Large\bfseries}{\thesection}{0.7em}{}
\titleformat{\subsection}{\sffamily\large\bfseries}{\thesubsection}{0.7em}{}
\titleformat{\subsubsection}{\sffamily\normalsize\bfseries}{\thesubsubsection}{0.7em}{}
\titlespacing*{\section}{0pt}{1.6ex plus .4ex minus .2ex}{0.9ex}
\titlespacing*{\subsection}{0pt}{1.3ex plus .3ex minus .2ex}{0.6ex}

\fancypagestyle{firstpage}{\fancyhf{}\fancyfoot[C]{\sffamily\footnotesize\thepage}}

\newcommand{\Rea}{\operatorname{Re}\varepsilon}
\newcommand{\Ima}{\operatorname{Im}\varepsilon}
\newcommand{\dE}{\Delta E^{\ddagger}}
\newcommand{\ab}{\alpha/\beta}

\begin{document}
\thispagestyle{firstpage}

\begin{center}
{\sffamily\bfseries\LARGE Machine learning interatomic potentials\\[2pt] and orientational defects in polymers\par}
\vspace{1.1em}
{\large
\mbox{Arnold Kiv\textsuperscript{1,2,*}},
\mbox{Volodymyr Soloviev\textsuperscript{1,3,*}},
\mbox{Andrii Tuzhykov\textsuperscript{1,*}},
\mbox{Andrii Bielinskyi\textsuperscript{4,5}},
\mbox{Taras Kavetskyy\textsuperscript{1,6,7}},
\mbox{Nataliia Hoivanovych\textsuperscript{6}},
\mbox{Ondrej Šauša\textsuperscript{7,8}},
\mbox{Jaroslav Rusnák\textsuperscript{7}},
\mbox{Helena Švajdlenková\textsuperscript{8,9}},
\mbox{Jolita Ostrauskaite\textsuperscript{10}},
and \mbox{Mykola Slusarenko\textsuperscript{3}}\par}
\vspace{0.9em}
{\small\itshape
\textsuperscript{1}South Ukrainian National Pedagogical University named after K.\,D.~Ushynsky, Odesa, Ukraine\\
\textsuperscript{2}Ben-Gurion University of the Negev, Beer-Sheva, Israel\\
\textsuperscript{3}Kryvyi Rih State Pedagogical University, Kryvyi Rih, Ukraine\\
\textsuperscript{4}State University of Economics and Technology, Kryvyi Rih, Ukraine\\
\textsuperscript{5}Kyiv National Economic University named after Vadym Hetman, Kyiv, Ukraine\\
\textsuperscript{6}Drohobych Ivan Franko State Pedagogical University, Drohobych, Ukraine\\
\textsuperscript{7}Institute of Physics, Slovak Academy of Sciences, Bratislava, Slovakia\\
\textsuperscript{8}Department of Nuclear Chemistry, Faculty of Natural Sciences, Comenius University, Bratislava, Slovakia\\
\textsuperscript{9}Polymer Institute, Slovak Academy of Sciences, Bratislava, Slovakia\\
\textsuperscript{10}Department of Polymer Chemistry and Technology, Kaunas University of Technology, Kaunas, Lithuania\par}
\vspace{0.6em}
{\small \textsuperscript{*}Corresponding authors:
A.~Kiv, \href{mailto:kivarnold20@gmail.com}{\nolinkurl{kivarnold20@gmail.com}};
V.~Soloviev, \href{mailto:vnsoloviev2016@gmail.com}{\nolinkurl{vnsoloviev2016@gmail.com}};
A.~Tuzhykov, \href{mailto:andrewtuzhykov@gmail.com}{\nolinkurl{andrewtuzhykov@gmail.com}}\par}
\vspace{0.4em}
{\small\sffamily November 2026\par}
\end{center}

\begingroup
\let\thefootnote\relax
\footnotetext{\textbf{Note.} The final version of this work was published as Chapter~6 of
P.~Petkov, M.\,E.~Achour, C.~Popov (eds.), \emph{Nanotechnological Advances in Environmental, Cyber and CBRN Security},
NATO Science for Peace and Security Series~B: Physics and Biophysics, Springer, Dordrecht, 2025, pp.~91--104,
\url{https://doi.org/10.1007/978-94-024-2316-7_6}.}
\endgroup

\vspace{0.4em}
\noindent\rule{\textwidth}{0.4pt}
\begin{abstract}
\noindent
Photopolymerization of acrylated epoxidized soybean oil (AESO) is a convenient model of a self-organizing process in a soft,
bio-based material: ultraviolet light starts a free-radical chain reaction, the growing network releases heat, and the
mobility of molecular groups changes on several time scales at once. We follow this process from two sides. Experimentally,
the real and imaginary parts of the dielectric permittivity of an AESO:VDM (1:0.5) mixture with 3~mol\% of the photoinitiator
DMPA were recorded at \SIrange{500}{5000}{Hz} under an intermittent on/off illumination protocol, and the resulting time series
were analysed with a sliding-window Shannon entropy and the Grassberger--Procaccia correlation integral. Both measures show that
short light pulses at the start of curing trigger strongly correlated, collective relaxation, which fades as the network forms,
leaving only localized reorientations active. We describe these reorientations as orientational defects and quantify their
energetics with the MACE-OFF23 machine learning interatomic potential, using relaxed dihedral scans of five rotational modes
($\alpha$, $\ab$, $\beta$, $\gamma$, $\delta$) of an AESO fragment. Compared with density functional theory (DFT) reference
profiles, MACE reproduces the shape of all five profiles (Pearson $r=0.95$--$0.98$) and the hierarchy of barriers---from
\SI{2.9}{eV} for the constrained main-chain rotation, through \SIrange{0.54}{0.81}{eV} for the acrylate rotations, to
\SIrange{0.16}{0.18}{eV} for the methyl and \SI{0.06}{eV} for the hydroxyl rotation---while underestimating the barrier heights by
3--20\%. A simple Arrhenius estimate shows that the acrylate rotations are the modes closest to the frequency window of the
dielectric measurement.

\end{abstract}
\noindent\rule{\textwidth}{0.4pt}

\section{Introduction}\label{sec:intro}

Polymers are a natural example of complex systems. Their macromolecules are organized hierarchically---from chemical bonds
and side groups, through chain segments, to a three-dimensional network---and their macroscopic properties emerge from the
interplay of these levels. Photopolymerization makes this complexity directly observable. Under illumination, a photoinitiator
produces free radicals, monomers add to growing chains, and within tens of seconds a liquid mixture turns into a cross-linked
solid. The same process changes the mobility of every polar group in the material, and therefore its response to an applied
electric field~\cite{woodward2021,romanini2023,moon2021}.

Bio-based monomers such as acrylated epoxidized soybean oil (AESO) are attractive for such studies. AESO is derived from a
renewable feedstock, cures rapidly under ultraviolet (UV) light, and its networks with the vanillin-derived comonomer vanillin
dimethacrylate (VDM) have been characterized in detail with respect to local free volume and diffusion properties~\cite{kralovic2024}.
At the same time, the molecule carries several groups that can reorient---acrylate, epoxy and hydroxyl groups attached to an
aliphatic main chain---and each of them contributes to the dielectric response in its own way.

Two complementary routes lead from such a material to an understanding of its dynamics. The first works with macroscopic
observables. Temperature, permittivity and other quantities recorded during curing form time series whose fluctuations carry
information on the internal transformations of the system; the methods of complex-systems theory make it possible to extract this
information and to recognize characteristic states and transitions~\cite{kiv2023dyn,kiv2023itest,kralovic2024}. The second
route is microscopic: quantum-mechanical calculations give the energies of molecular conformations and, in particular, the barriers
that a reorienting group has to overcome. Such calculations are, however, expensive, which limits the size of the systems and the
number of configurations that can be studied. Machine learning interatomic potentials (MLIPs) remove much of this limitation: they
are trained on first-principles data and reproduce first-principles energies and forces at a small fraction of the
cost~\cite{chong2024,batatia2022mace}.

In this work we combine the two routes. The dielectric relaxation of AESO during photopolymerization, with and without a
photoinitiator, was studied for the first time under a protocol in which the light is switched on and off with variable pulse
and pause lengths. The protocol makes it possible to observe the relatively fast photopolymerization (tens of seconds) and the
relatively slow dielectric relaxation within one experiment~\cite{kralovic2024}. The microscopic interpretation of the measured
response is then based on the concept of orientational defects, whose energetics we compute with MLIPs of the MACE
family~\cite{batatia2022mace,batatia2025mp,kovacs2025off}. The specific contributions of the paper are:
\begin{enumerate}[label=(\roman*),leftmargin=2.2em,itemsep=1pt,topsep=3pt]
  \item a sliding-window entropy and correlation analysis of the permittivity time series that separates collective and localized
        relaxation stages of curing (\cref{sec:complexity});
  \item a classification of the mobile groups of AESO as orientational defects and their assignment to the $\alpha$, $\ab$,
        $\beta$, $\gamma$ and $\delta$ relaxation modes (\cref{sec:od});
  \item a reproducible MLIP-based pipeline for relaxed rotational scans built on the Atomic Simulation Environment (\cref{sec:methods});
  \item a quantitative comparison of the MACE energy profiles with DFT reference profiles and an order-of-magnitude link between
        the computed barriers and the frequency window of the measurement (\cref{sec:results}).
\end{enumerate}

\section{Materials and experimental protocol}\label{sec:experiment}

\subsection{Materials}

\begin{figure}[t]
  \centering
  \begin{subfigure}[b]{0.47\textwidth}\centering
    \includegraphics[width=\linewidth]{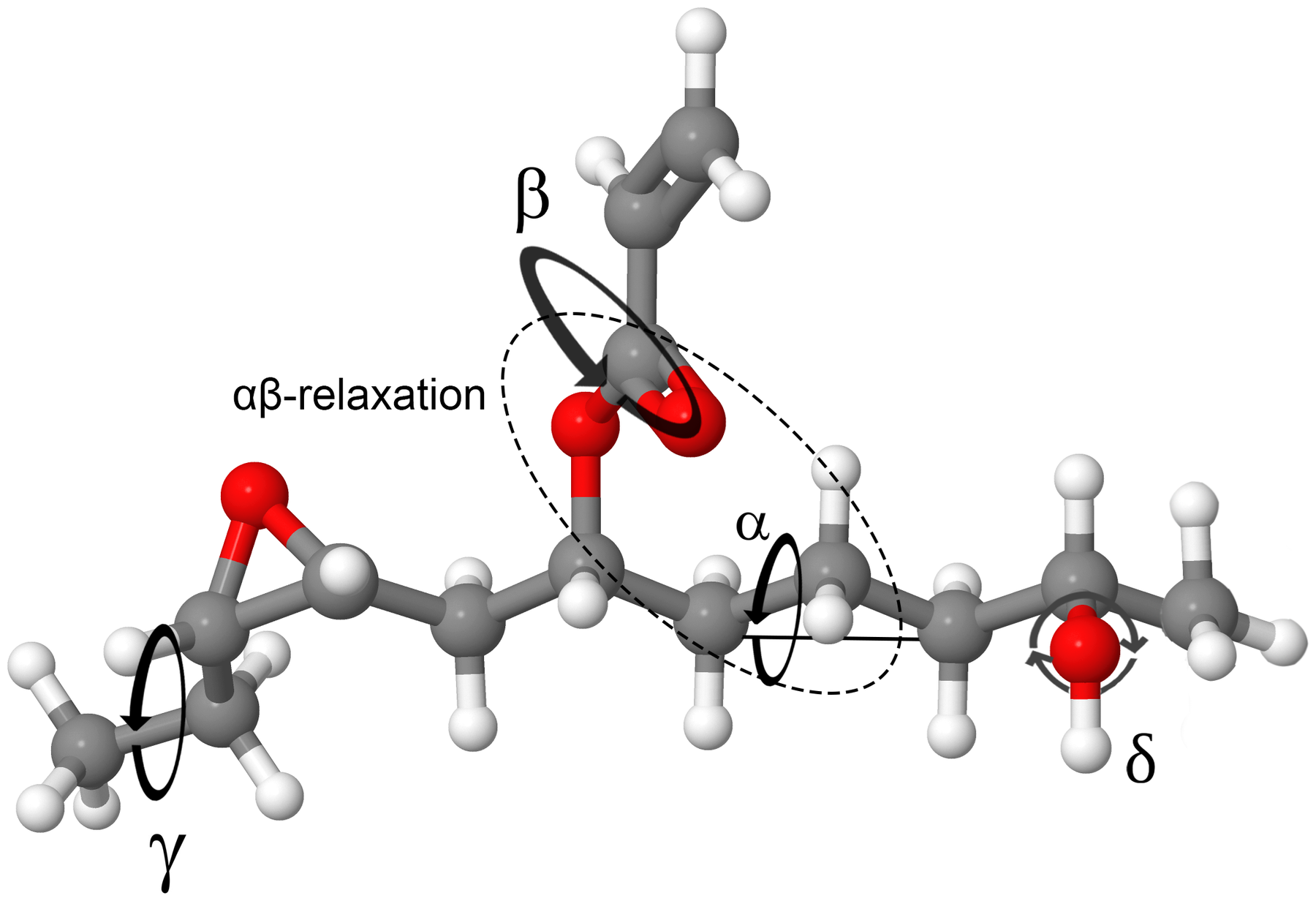}\caption{}\end{subfigure}\hfill
  \begin{subfigure}[b]{0.255\textwidth}\centering
    \includegraphics[width=\linewidth]{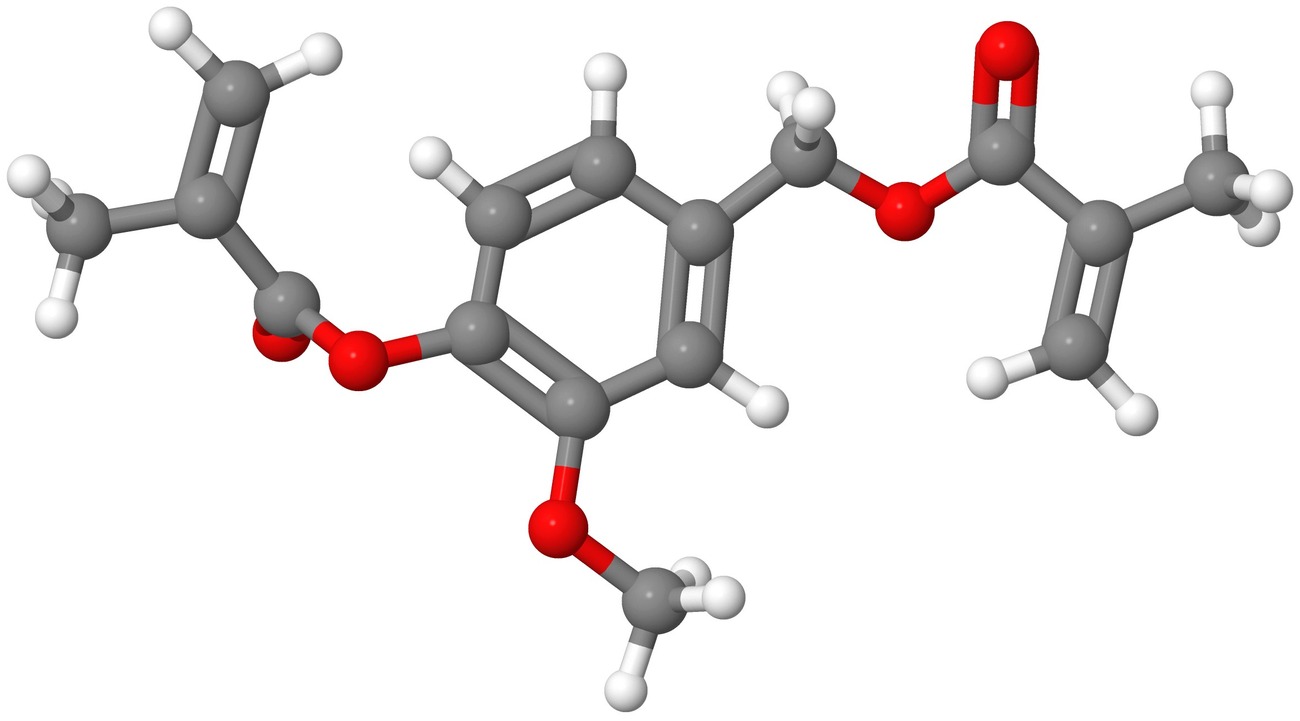}\caption{}\end{subfigure}\hfill
  \begin{subfigure}[b]{0.255\textwidth}\centering
    \includegraphics[width=\linewidth]{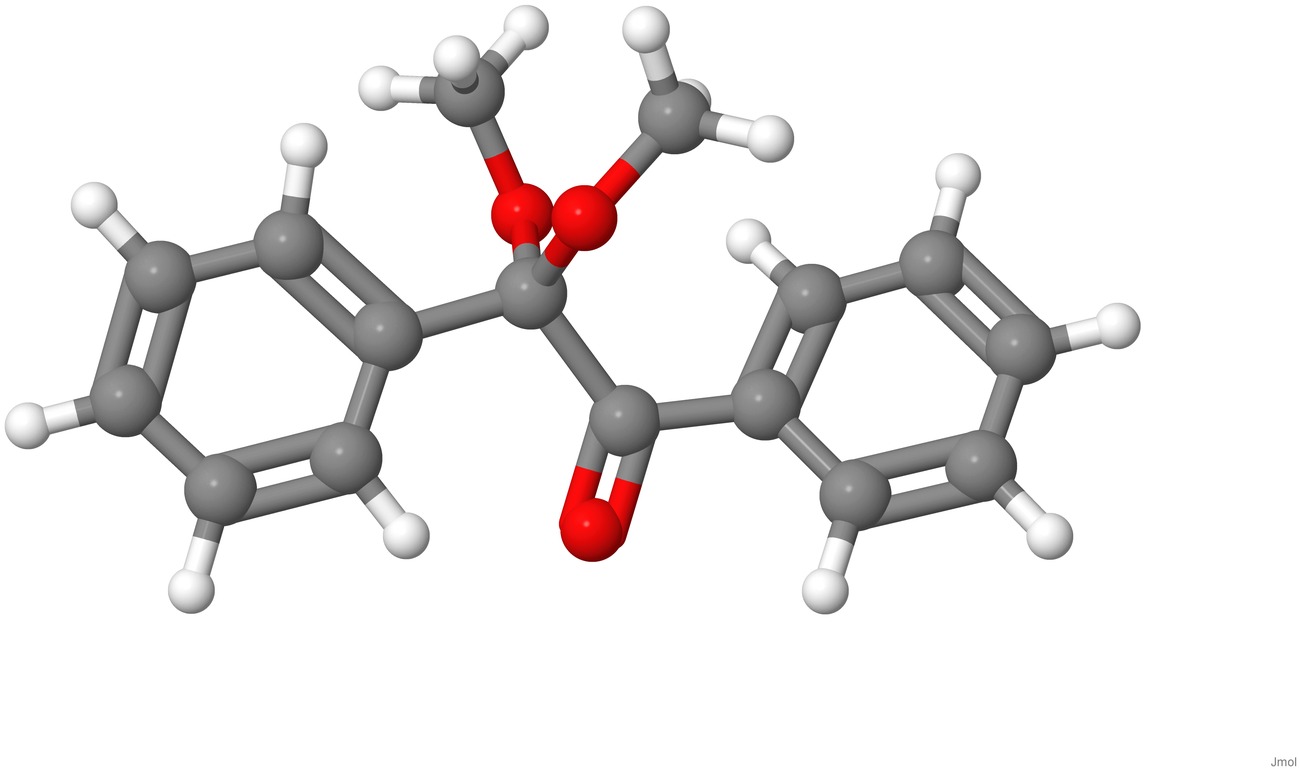}\caption{}\end{subfigure}
  \caption{Components of the photocurable mixture. (a) Fragment of acrylated epoxidized soybean oil (AESO) with the main chain
  and the acrylate, hydroxyl and epoxy side groups; arrows mark the rotations associated with the $\alpha$, $\ab$, $\beta$,
  $\gamma$ and $\delta$ relaxation modes (\cref{tab:modes}). (b) Vanillin dimethacrylate (VDM). (c) The photoinitiator
  2,2-dimethoxy-2-phenylacetophenone (DMPA). Carbon is grey, oxygen red, hydrogen white.}
  \label{fig:molecules}
\end{figure}

The studied system consists of AESO, the comonomer VDM, and the photoinitiator (PI) 2,2-dimethoxy-2-phenylacetophenone (DMPA);
their structures are shown in \cref{fig:molecules}. AESO and DMPA were purchased from Sigma-Aldrich and VDM from Specific Polymers.
The molar ratio AESO:VDM was 1:0.5, and the PI concentration was 3~mol\% with respect to the total amount of monomers. Adding VDM
to AESO noticeably increases the viscosity of the mixture~\cite{kralovic2024}. \Cref{fig:molecules}a also marks the groups that are
expected to dominate the relaxation behaviour---the acrylate, epoxy and hydroxyl groups---and the rotations considered in
\cref{sec:od,sec:results}.

\subsection{Photocuring and dielectric measurements}

Photopolymerization was initiated with a Jaxman U1 LED source emitting at \SI{365}{nm}. The real ($\Rea$) and imaginary ($\Ima$)
parts of the dielectric permittivity were measured at room temperature at frequencies of \SIrange{500}{5000}{Hz}. Dielectric
relaxation spectroscopy (DRS) is particularly well suited to this task: it is sensitive to the microscopic dynamics of polymer
systems~\cite{woodward2021} and is one of the principal methods for studying their structure and molecular
mobility~\cite{romanini2023,kremer2003}.

The illumination followed an on/off protocol with three stages. In the first stage the light was switched on for \SI{1}{s} at a
time, with dark intervals of increasing length between the pulses. In the second stage the pulses were extended to \SI{10}{s},
with the same sequence of dark intervals. In the third stage the sample was illuminated continuously until the end of the
measurement; the whole record spans about \SI{8000}{s} (\cref{fig:complexity}). Samples with and without the photoinitiator were
studied; the results discussed below refer to AESO:VDM~1:0.5 with PI. The time series of $\Rea$ and $\Ima$ were analysed with the
quantitative measures of complexity described in the next section~\cite{kiv2023dyn,kiv2023itest,shannon1948,grassberger1983}.

\section{Complexity analysis of the dielectric response}\label{sec:complexity}

Photopolymerization is a dynamic, synergetic process in which chemical reactions, heat release and structural rearrangements are
coupled. A suitable general toolkit for such processes is provided by the theory of complex systems~\cite{kiv2023dyn,kiv2023itest},
in particular by entropy-based~\cite{shannon1948} and correlation-based~\cite{grassberger1983} measures of complexity. In both cases
the measure is computed for a short fragment (window) of the time series; the window is then shifted by a fixed step and the
calculation is repeated. Characteristic changes in the resulting sequence of values, interpreted according to the nature of
each measure, serve as indicators of structural and dynamical transformations in the system~\cite{kiv2023dyn,kiv2023itest}.

\subsection{Sliding-window procedure}

Let $\{x_t\}_{t=1}^{T}$ be the recorded series (here $\Rea$ or $\Ima$ at a fixed frequency). For a window of length $w$ shifted with
a step $s$, the $k$-th fragment is
\begin{equation}
  X^{(k)} = \bigl\{x_{(k-1)s+1},\,x_{(k-1)s+2},\,\ldots,\,x_{(k-1)s+w}\bigr\},\qquad k=1,2,\ldots,\Bigl\lfloor\frac{T-w}{s}\Bigr\rfloor+1 .
  \label{eq:window}
\end{equation}
Each measure is evaluated on $X^{(k)}$ and assigned to the end of the window, so that it depends only on past data. In the
analysis shown in \cref{fig:complexity} the window covers \SI{250}{s} and the step is \SI{1}{s}.

\subsection{Shannon entropy}

Entropy measures the average amount of information carried by each symbol of a process~\cite{bielinskyi2021}. For a stationary
random process $X=\{x_1,x_2,\ldots,x_N\}$ with probability distribution $p(\cdot)$ the information (Shannon) entropy (ShEn)
is~\cite{shannon1948}
\begin{equation}
  H(X) = -\sum_{x\in X} p(x)\,\log p(x).
  \label{eq:shen}
\end{equation}
In practice, $p(x)$ is estimated from the empirical distribution of the values inside the window.
ShEn quantifies the uncertainty of the source: it is maximal for an irregular, unpredictable signal and tends to zero when the
signal is highly regular. A drop of ShEn inside a window therefore signals that the dynamics has become more ordered.

\subsection{Correlation integral}

To build a model of the underlying dynamics, one also needs information on the correlation properties of its attractor. Besides
the entropy approach, these can be studied with the correlation integral introduced by Grassberger and
Procaccia~\cite{grassberger1983}. The scalar signal is first embedded in a pseudo-phase space by the method of
delays~\cite{takens1981},
\begin{equation}
  \vec{y}_i = \bigl(x_i,\,x_{i+\tau},\,\ldots,\,x_{i+(m-1)\tau}\bigr),
  \label{eq:embedding}
\end{equation}
with embedding dimension $m$ and delay $\tau$. The correlation integral is the fraction of pairs of phase-space points that are
closer to each other than a distance $r$,
\begin{equation}
  C(r) = \lim_{N\to\infty}\frac{1}{N^2}\sum_{\substack{i,j=1\\ i\neq j}}^{N}\Theta\bigl(r-\lVert\vec{y}_i-\vec{y}_j\rVert\bigr),
  \label{eq:corrint}
\end{equation}
where $N$ is the number of phase-space vectors, $\Theta(\cdot)$ is the Heaviside step function, and $r$ is the radius of a
hypersphere in the pseudo-phase space. For a finite window the limit is replaced by the corresponding sum over the $N$ available
vectors. The larger $C(r)$, the more densely the phase-space points are packed inside spheres of radius $r$, i.e., the stronger the
self-organization of the process.

\subsection{Dynamics of the dielectric response during curing}

\begin{figure}[t]
  \centering
  \includegraphics[width=0.86\textwidth]{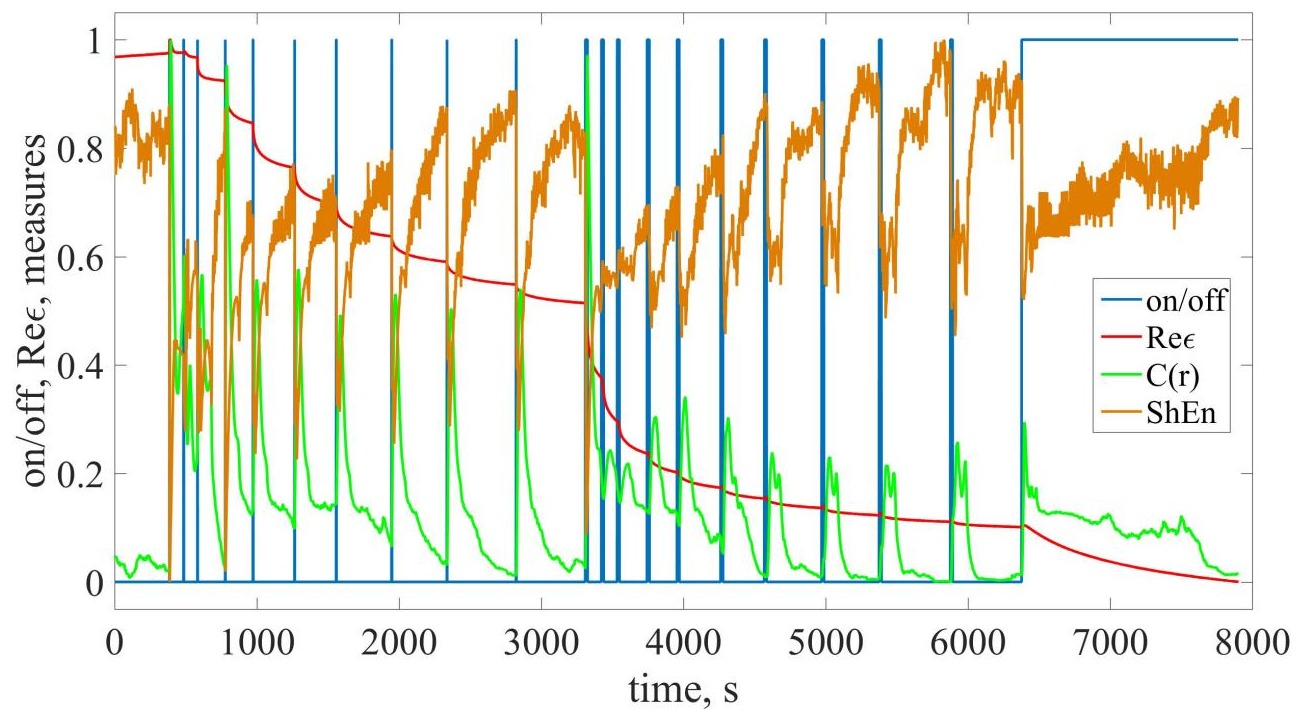}
  \caption{Evolution of $\Rea$, the correlation integral $C(r)$ and the Shannon entropy (ShEn) during photocuring of AESO:VDM~1:0.5
  with PI. $C(r)$ and ShEn were calculated in a sliding window of \SI{250}{s} shifted by \SI{1}{s}. The on/off trace shows the
  illumination protocol: \SI{1}{s} pulses separated by increasing dark intervals, then \SI{10}{s} pulses with the same intervals,
  and finally continuous illumination. All curves are scaled to the range $[0,1]$.}
  \label{fig:complexity}
\end{figure}

\Cref{fig:complexity} shows a typical result of the analysis for the real part of the permittivity. Three features stand out.

First, during the stage of short pulses, each switching-on of the light is followed by a sharp rise of $C(r)$ and a drop of ShEn.
Both measures then relax back over a few hundred seconds, so that ShEn develops a characteristic saw-tooth pattern between
consecutive pulses. The simultaneous increase of correlation and decrease of entropy indicates highly correlated processes that
involve the cooperative motion of a large number of atoms---in our interpretation, the collective, self-organized reorientation of
orientational defects (\cref{sec:od}).

Second, every pulse starts new free-radical reactions, which stimulate both the correlated and other, less correlated processes.
As curing proceeds, however, the response to each pulse weakens. The amplitude of the $C(r)$ bursts decreases, and under
continuous illumination the strongly correlated relaxation fades away almost completely: $C(r)$ settles at a low level and ShEn
returns to high values. Only localized orientational defects remain active at this stage.

Third, $\Rea$ decreases throughout the experiment, with distinct steps when the light is switched on, most clearly at the first
\SI{10}{s} pulse. This behaviour is consistent with the progressive restriction of dipolar reorientation as the polymer network
forms.

Such dynamics reflects the changing population and mobility of defects under irradiation and during network formation. The
structure of the reacting components allows us to identify the main mobile structural units, most of which are orientational in
nature. In the next section we therefore introduce the concept of orientational defects and the quantities that characterize them.

\section{Orientational defects}\label{sec:od}

\subsection{The concept}

Polymers are used in fields ranging from biomedicine to electronics, and their properties depend strongly on molecular structure
and on the defects it contains~\cite{jangizehi2020,mishra2024}. Among these, orientational defects (ODs) play a special role in
structural transformations and functional behaviour. An OD is a local configuration in which an atomic or molecular group is rotated
away from its regular orientation; it is associated with a local redistribution of electron density, and its creation, motion and
annihilation proceed through rotations of groups within the matrix. Knowledge of the energy landscape of these rotations is
therefore essential for designing polymeric materials for specific applications.

\begin{figure}[t]
  \centering
  \begin{subfigure}[b]{0.50\textwidth}\centering
    \includegraphics[width=\linewidth]{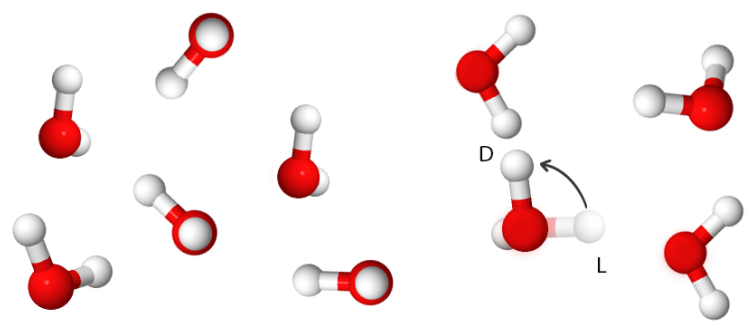}\caption{}\end{subfigure}\hfill
  \begin{subfigure}[b]{0.46\textwidth}\centering
    \includegraphics[width=\linewidth]{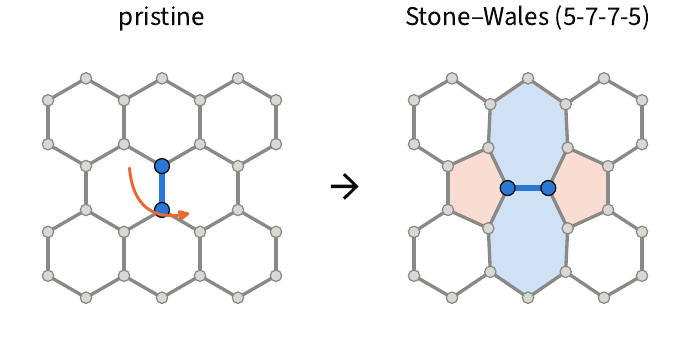}\caption{}\end{subfigure}\\[4pt]
  \begin{subfigure}[b]{0.31\textwidth}\centering
    \includegraphics[width=\linewidth]{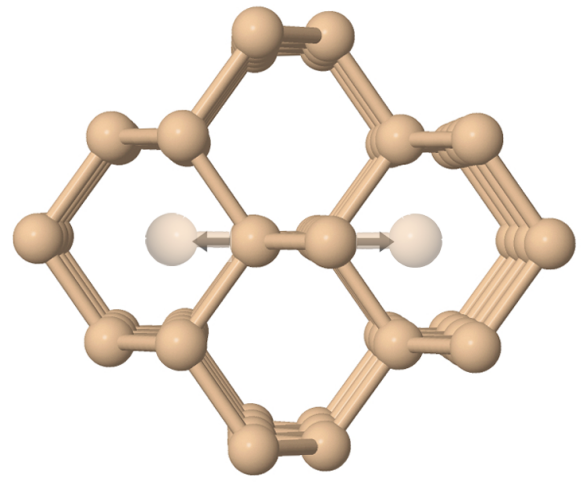}\caption{}\end{subfigure}\hspace{0.08\textwidth}
  \begin{subfigure}[b]{0.38\textwidth}\centering
    \includegraphics[width=\linewidth]{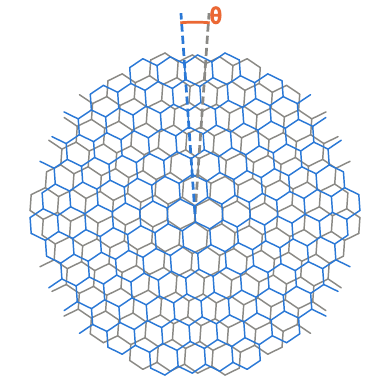}\caption{}\end{subfigure}
  \caption{Orientational defects in different materials. (a) Bjerrum D and L defects in ice, created by the rotation of a water
  molecule~\cite{bjerrum1952,dekoning2006}. (b) Stone--Wales defect in graphene: a $90^\circ$ rotation of one C--C bond (blue)
  turns four hexagons into two pentagons and two heptagons~\cite{basak2021}. (c) Divacancy with ``rotated bonds'' in the diamond
  lattice of silicon~\cite{kiv1979}. (d) Twist defect: two graphene layers rotated with respect to each other by an angle
  $\theta$~\cite{popov2023}. Panels (b) and (d) are schematic drawings made for this preprint.}
  \label{fig:od}
\end{figure}

Historically, the term ``orientational defect'' was introduced by Bjerrum in the study of ice~\cite{bjerrum1952}
(\cref{fig:od}a). A Bjerrum pair is produced when a water molecule rotates so that one hydrogen bond receives two protons
(a D defect) and another receives none (an L defect). Subsequent thermally activated rotations of neighbouring molecules can
separate the D and L defects and finally make them independent. As a free defect moves through the crystal, it reorients the
molecules along its path, which provides a molecular mechanism of electric polarization and conduction in
ice~\cite{dekoning2006}.

In diamond-like semiconductors, metastable OD configurations with ``rotated bonds'' were first considered by Kiv and
Soloviev~\cite{kiv1979} to interpret low-temperature data on the formation and migration of point defects in covalent crystals.
\Cref{fig:od}c shows such a divacancy: the atoms that left the vacant sites do not go far away, as in a classical divacancy, but
stay close to the neighbouring sites. In other words, the configuration consists of two unseparated Frenkel pairs. Several defect
states of graphene are orientational as well. The Stone--Wales defect changes the connectivity of two $\pi$-bonded carbon atoms by
rotating their bond by $90^\circ$ about its midpoint (\cref{fig:od}b)~\cite{basak2021}, and a twist defect is an orientational
displacement of one graphene layer relative to another (\cref{fig:od}d)~\cite{popov2023}. Some structural configurations of
amorphous semiconductors, in particular amorphous silicon~\cite{morrow2024}, can probably be regarded as orientational defects too.
In polymers and nanocomposites, however, ODs are clearly the dominant defect type. They govern most electronic and atomic processes,
including relaxation, and control in particular degradation processes~\cite{nie2016}.

\subsection{Orientational defects in AESO and the relaxation hierarchy}

Dielectric relaxation in polymers is a superposition of several competing, thermally activated processes, conventionally labelled
$\alpha$, $\beta$, $\gamma$, \ldots~\cite{kremer2003}. The primary $\alpha$-relaxation results from the reorientation of chain
segments (main chains), involves a large group of atoms and is distinctly collective. The secondary $\beta$-, $\gamma$-, \ldots
relaxations arise from the motion of side groups attached to the chains. Among them, an $\ab$ process is sometimes distinguished,
in which the side group is most strongly coupled to the main chain.

For AESO (\cref{fig:molecules}a) we assign these modes to specific rotations, summarized in \cref{tab:modes}. The low-frequency
$\alpha$-relaxation corresponds to rotation of a main-chain segment, i.e., to segmental or crankshaft motion. The $\ab$ and $\beta$
processes are caused by rotation of the acrylate functional group about the C--O bond that connects it to the main chain; they
appear mainly at intermediate frequencies and contribute strongly to dielectric loss. At still higher frequencies AESO can show
$\gamma$- and further relaxations related to the rotation of methyl (CH$_3$) and hydroxyl (C--O--H) groups bonded to the main
chain. Each of these rotations creates and removes an orientational defect, and its energy barrier is the quantity that we compute
in \cref{sec:results}.

\begin{table}[t]
  \centering
  \caption{Rotational modes of the AESO fragment considered in this work (see the arrows in \cref{fig:molecules}a) and the angular
  step $\Delta\varphi_{\mathrm{step}}$ of the corresponding relaxed scan.}
  \label{tab:modes}
  \small
  \begin{tabular}{@{}l>{\raggedright\arraybackslash}p{0.5\textwidth}>{\raggedright\arraybackslash}p{0.17\textwidth}c@{}}
    \toprule
    Mode & Rotating unit & Character & $\Delta\varphi_{\mathrm{step}}$ \\
    \midrule
    $\alpha$  & main-chain segment, rotated about the chain axis & segmental, collective & $5^\circ$ \\
    $\ab$     & acrylate side group together with its C--O link to the main chain & side group coupled to the chain & $5^\circ$ \\
    $\beta$   & acrylate group about the C--O bond to the main chain & local side group & $5^\circ$ \\
    $\gamma$  & methyl group & local, small & $1^\circ$ \\
    $\delta$  & hydroxyl group (O--H about the C--O bond) & local, small & $1^\circ$ \\
    \bottomrule
  \end{tabular}
\end{table}

Primary and secondary relaxations are characterized by their own activation energies, which can be estimated by quantum-mechanical
methods. Standard DFT calculations of energies and structures, however, require considerable computing resources. Recent progress
in machine learning for materials science~\cite{chong2024} allows this task to be reduced to the construction---or simply the
use---of machine learning interatomic potentials, which yield the required parameters with good accuracy at a limited cost. MLIPs
predict potential energy surfaces and reaction paths of the kind associated with ODs; by combining the accuracy of ab~initio
reference data with the efficiency of machine learning, they make it possible to explore complex chemical systems at scales that
are not accessible to direct first-principles calculations.

\section{Computational methodology}\label{sec:methods}

\subsection{Machine learning interatomic potentials and the MACE architecture}

Our methodology rests on MLIPs of the MACE family~\cite{batatia2022mace}, which we use to predict the potential energy surfaces and
forces associated with ODs in AESO. Like most MLIPs, MACE writes the total energy of a configuration of atoms with positions
$\{\vec{r}_i\}$ and chemical species $\{Z_i\}$ as a sum of atomic contributions, each depending only on the neighbourhood of the atom
within a cutoff radius $r_{\mathrm{c}}$, and obtains the forces as exact derivatives of this energy:
\begin{equation}
  E = \sum_{i} E_i\bigl(\{(\vec{r}_{ij},Z_j)\}_{r_{ij}<r_{\mathrm{c}}}\bigr),\qquad
  \vec{F}_i = -\frac{\partial E}{\partial \vec{r}_i}.
  \label{eq:energy}
\end{equation}
The atomic energies are produced by an equivariant message-passing neural network. In each layer, information from neighbours
is first collected into atomic features built from learnable radial functions and spherical harmonics of the bond directions,
\begin{equation}
  A_{i} = \sum_{j\in\mathcal{N}(i)} R(r_{ij})\,Y(\hat{r}_{ij})\,h_j ,
  \label{eq:ace}
\end{equation}
where $h_j$ are the current features of atom $j$ (schematic notation, channel and angular indices omitted). These features are
then combined into symmetric products of order up to $\nu$, which gives messages of body order $\nu+1$---in the original
architecture four-body messages---so that two message-passing layers are sufficient for high accuracy~\cite{batatia2022mace}.
Because of the equivariant construction, the predicted energy is invariant, and the forces are equivariant, with respect to
rotations, translations and permutations of identical atoms.

During prediction, a trained MACE model takes an atomic configuration and returns the total potential energy and the forces on all
atoms. These are exactly the quantities needed for the structural optimizations and molecular dynamics runs described below.

\subsection{Foundation models: MACE-MP-0 and MACE-OFF23}

We employed pretrained (``foundation'') MACE models, which can be used without system-specific training. MACE-MP-0~\cite{batatia2025mp}
covers 89 chemical elements and was trained on inorganic crystal data from the Materials Project computed with the PBE(+U) functional;
it is intended as a general-purpose model for materials chemistry. MACE-OFF23~\cite{kovacs2025off} is a series of transferable force
fields for neutral organic molecules composed of ten elements (H, C, N, O, F, P, S, Cl, Br and I). It was trained on the SPICE
dataset~\cite{eastman2023spice}, i.e., on energies and forces computed at the $\omega$B97M-D3(BJ)/def2-TZVPPD level, and is available
in small (S), medium (M) and large (L) variants that trade accuracy against speed.

MACE-OFF23 is the natural choice for AESO, a bio-based molecule containing only H, C and O. Of particular importance here is its
ability to reproduce potential energy surfaces and barrier heights of dihedral rotations, which was demonstrated on the
TorsionNet-500 and biaryl torsion benchmarks~\cite{kovacs2025off}. This is precisely the information needed to describe the
conformational energy landscapes associated with ODs in AESO during photopolymerization. The S and M variants combine high accuracy
with a computational cost that permits extensive sampling of configuration space, larger fragments and longer time scales. The
pipeline shown in \cref{fig:pipeline} uses the medium model, MACE-OFF23(M), evaluated on a GPU.

\subsection{Atomic Simulation Environment, constraints and optimizers}

The setup, manipulation and analysis of the simulations were carried out with the Atomic Simulation Environment
(ASE)~\cite{larsen2017}, a Python library that provides a common interface to structures, constraints, optimizers and
calculators, including MACE. The AESO fragment is placed in a periodic simulation cell, and ASE handles the periodic images and
boundary conditions consistently throughout the simulations.

ASE also provides the constraints needed for a rotational scan. We use the \texttt{FixAtoms} constraint to keep selected atoms of the
AESO fragment fixed. This preserves the overall molecular framework while the sub-molecule of interest rotates, so that the scan
probes the energy landscape of a specific OD without introducing unwanted distortions elsewhere in the molecule.

For structural relaxation we use the Fast Inertial Relaxation Engine (FIRE)~\cite{bitzek2006,guenole2020}. FIRE propagates damped
molecular dynamics in which the velocity is continuously mixed towards the direction of the force,
\begin{equation}
  \vec{v} \leftarrow (1-a)\,\vec{v} + a\,\lvert\vec{v}\rvert\,\hat{\vec{F}},
  \label{eq:fire}
\end{equation}
increases the time step and decreases the mixing parameter $a$ while the power $P=\vec{F}\cdot\vec{v}$ stays positive, and stops
and restarts the motion as soon as $P$ becomes negative. This combination of inertia and gradient information makes FIRE robust and
efficient for finding the nearest local minimum. As an alternative, Langevin molecular dynamics,
\begin{equation}
  m_i\ddot{\vec{r}}_i = \vec{F}_i - \gamma_{\mathrm{L}}\,m_i\dot{\vec{r}}_i + \vec{\xi}_i(t),\qquad
  \langle \xi_{i\mu}(t)\,\xi_{j\nu}(t')\rangle = 2\gamma_{\mathrm{L}} m_i k_{\mathrm{B}}T\,\delta_{ij}\delta_{\mu\nu}\delta(t-t'),
  \label{eq:langevin}
\end{equation}
can be used; its friction and random forces mimic a heat bath and allow a more realistic, thermally driven exploration of the
potential energy surface.

\subsection{Simulation pipeline}

\begin{figure}[t]
  \centering
  \includegraphics[width=\textwidth]{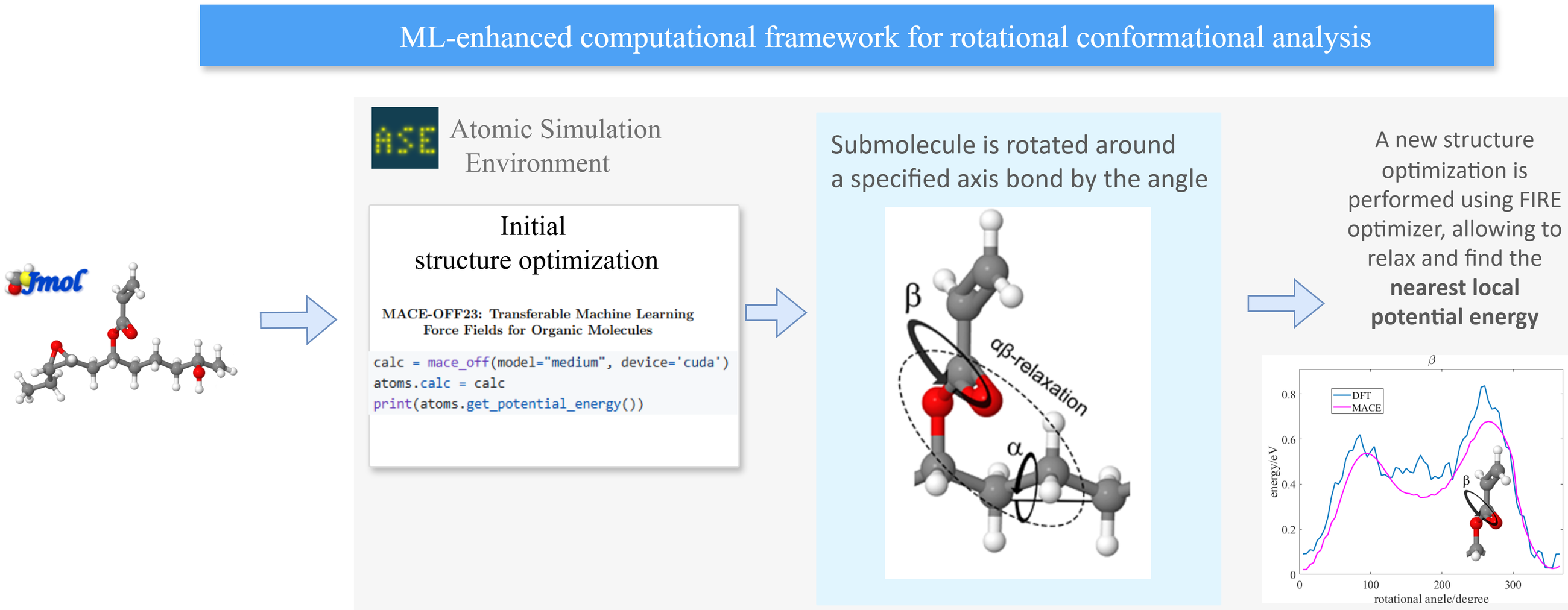}
  \caption{MLIP-based pipeline for the rotational conformational analysis. The structure is loaded into ASE and pre-optimized with
  MACE-OFF23; a sub-molecule is then rotated step by step about a selected bond, each rotated structure is relaxed with FIRE to the
  nearest local minimum, and the energies are collected into a rotational energy profile.}
  \label{fig:pipeline}
\end{figure}

The pipeline (\cref{fig:pipeline,alg:scan}) explores the conformational space systematically, locates local energy minima and
yields the relaxation energy profile associated with each OD.

\emph{Initialization.} The AESO structure is generated with molecular modelling tools (or taken from experimental data) and read
into ASE from a standard format such as XYZ or PDB. It is then relaxed to a local energy minimum, either with FIRE, which minimizes
the potential energy directly, or with Langevin dynamics~\eqref{eq:langevin}, depending on the required balance between accuracy and
cost.

\emph{Rotational scan.} The sub-molecule of interest is rotated about the chosen bond axis in increments $\Delta\varphi_{\mathrm{step}}$
(\cref{tab:modes}). After every increment the structure is relaxed again with FIRE (or Langevin dynamics), which lets the rotated
group and its surroundings settle into the nearest local minimum compatible with the imposed rotation. Two parameters control each
relaxation: the force threshold $f_{\max}$---the optimization is considered converged when the force on every free atom is below
$f_{\max}$---and the maximum number of steps, which prevents the algorithm from stalling when convergence is slow.
Because each step starts from the relaxed structure of the previous one, the scan follows a continuous path through configuration
space.

\emph{Collection and comparison.} The optimized structures and their energies are stored and assembled into the rotational
energy profile $E(\Delta\varphi)$. Comparing these profiles with reference calculations and with experiment validates the accuracy
and the predictive power of the MLIP approach.

\begin{algorithm}[t]
  \caption{Relaxed rotational scan of an orientational defect with an MLIP}
  \label{alg:scan}
  \small
  \begin{algorithmic}[1]
    \Require structure $\mathcal{S}_0$; rotation axis (bond $j$--$k$); set $M$ of atoms of the rotating sub-molecule; fixed set $F$;
             step $\Delta\varphi_{\mathrm{step}}$; force threshold $f_{\max}$; step limit $n_{\max}$
    \State attach the MACE calculator to $\mathcal{S}_0$
    \State $\mathcal{S}\gets\textsc{Relax}(\mathcal{S}_0)$ \Comment{FIRE or Langevin dynamics}
    \State $E_0\gets E(\mathcal{S})$
    \For{$n=1,\ldots,360^\circ/\Delta\varphi_{\mathrm{step}}$}
      \State rotate the atoms in $M$ by $\Delta\varphi_{\mathrm{step}}$ about the axis $j\to k$
      \State apply \texttt{FixAtoms}$(F)$
      \State $\mathcal{S}\gets\textsc{FIRE}(\mathcal{S};\,f_{\max},\,n_{\max})$ \Comment{nearest local minimum}
      \State $E_n\gets E(\mathcal{S})$;\; store $\mathcal{S}$
    \EndFor
    \State \Return profile $\{(n\,\Delta\varphi_{\mathrm{step}},\,E_n-\min_m E_m)\}$
  \end{algorithmic}
\end{algorithm}

\subsection{Reference calculations and accuracy measures}

The MACE profiles were benchmarked against reference profiles obtained with density functional theory using the GPAW
code~\cite{mortensen2024gpaw} along the same rotational coordinates.
Each profile, DFT or MACE, is referred to its own minimum. The rotational barrier is defined as
\begin{equation}
  \dE = \max_{\varphi} E(\varphi) - \min_{\varphi} E(\varphi),
  \label{eq:barrier}
\end{equation}
and the agreement between the two profiles sampled at the same $n$ angles is measured by the mean absolute error, the root-mean-square
error and the Pearson correlation coefficient,
\begin{equation}
  \mathrm{MAE} = \frac{1}{n}\sum_{k=1}^{n}\bigl\lvert E^{\mathrm{MACE}}_k - E^{\mathrm{DFT}}_k\bigr\rvert,\qquad
  \mathrm{RMSE} = \Bigl[\frac{1}{n}\sum_{k=1}^{n}\bigl(E^{\mathrm{MACE}}_k - E^{\mathrm{DFT}}_k\bigr)^2\Bigr]^{1/2}.
  \label{eq:metrics}
\end{equation}

\section{Results and discussion}\label{sec:results}

\subsection{Rotational energy profiles}

\begin{figure}[tbp]
  \centering
  \includegraphics[width=\textwidth]{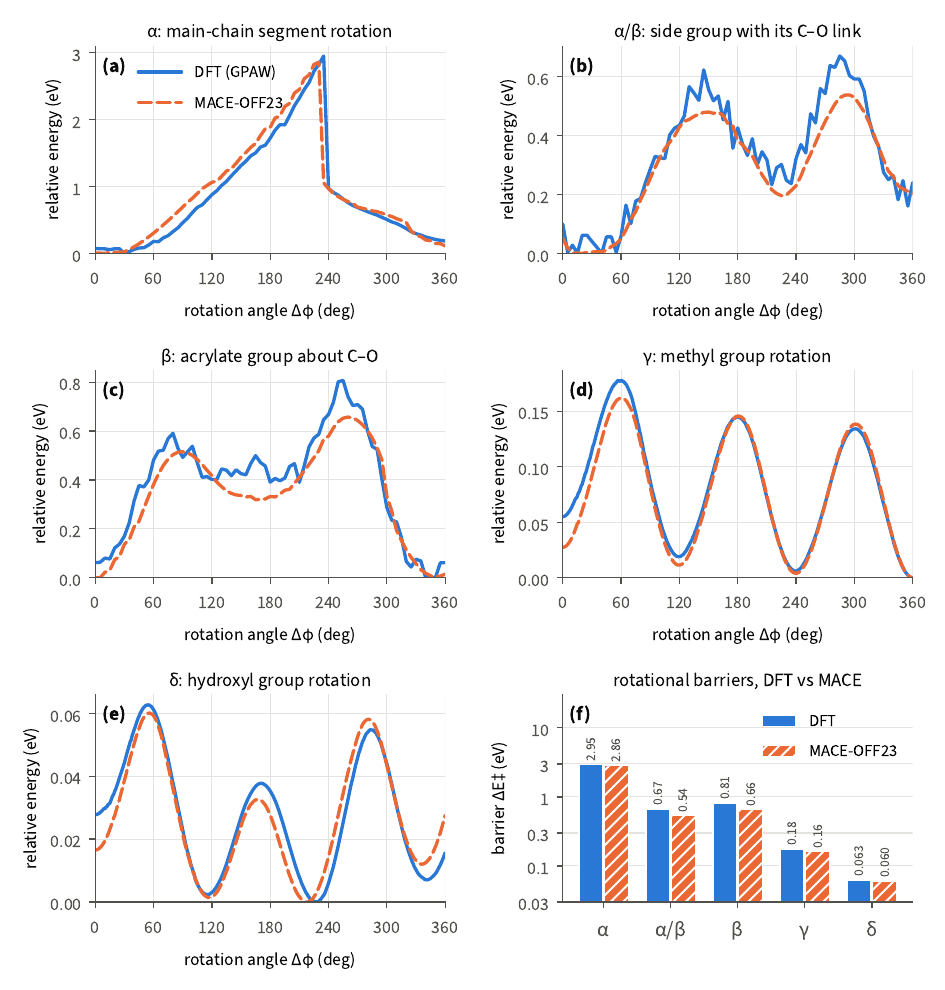}
  \caption{Rotational energy profiles of the AESO fragment for the modes of \cref{tab:modes} (see the arrows in
  \cref{fig:molecules}a): (a)~$\alpha$, (b)~$\ab$, (c)~$\beta$, (d)~$\gamma$, (e)~$\delta$. Solid lines: DFT reference (GPAW);
  dashed lines: MACE-OFF23 prediction. $\Delta\varphi$ is the rotation angle measured from the initial optimized conformation, and
  each profile is referred to its own minimum. (f)~Rotational barriers $\dE$ (\cref{eq:barrier}) from both methods on a logarithmic
  scale.}
  \label{fig:profiles}
\end{figure}

The five relaxed scans are shown in \cref{fig:profiles}. They differ by almost two orders of magnitude in energy and also in shape.

\emph{$\alpha$ mode} (\cref{fig:profiles}a). Rotating a main-chain segment while the rest of the framework is held fixed costs a
steadily increasing amount of energy, up to $\approx\SI{2.9}{eV}$ near $\Delta\varphi\approx230$--$235^\circ$. At the next step the
energy drops abruptly to $\approx\SI{1}{eV}$ and then decreases smoothly to its initial level. The drop indicates that the
constrained structure suddenly escapes into another conformational basin. The profile is thus strongly asymmetric, and the
maximum reflects the steric strain accumulated by a forced, essentially rigid rotation rather than a saddle point of a cooperative
process. MACE reproduces the shape of the curve, including the abrupt drop, which it places one step ($5^\circ$) earlier than DFT.

\emph{$\ab$ and $\beta$ modes} (\cref{fig:profiles}b,c). The rotations of the acrylate group give two-humped profiles with barriers
of \SIrange{0.5}{0.8}{eV}. For $\ab$ the maxima lie near $145$--$155^\circ$ and $285$--$295^\circ$, separated by a local minimum
at $\approx225$--$230^\circ$; for $\beta$ the main maximum lies near $255$--$260^\circ$, and a secondary one near
$80$--$90^\circ$ is followed by a broad plateau. The DFT profiles fluctuate noticeably from point to point, whereas the MACE profiles
are smooth.

\emph{$\gamma$ and $\delta$ modes} (\cref{fig:profiles}d,e). The rotations of the methyl and hydroxyl groups produce nearly threefold
periodic profiles with maxima at $\approx 60^\circ$, $180^\circ$ and $300^\circ$ ($\gamma$) and at $\approx 55^\circ$, $170^\circ$
and $285^\circ$ ($\delta$), as expected for rotors with three equivalent or nearly equivalent positions. The barriers are small:
\SIrange{0.13}{0.18}{eV} for $\gamma$ and \SIrange{0.04}{0.06}{eV} for $\delta$, depending on the maximum. For these modes MACE and
DFT agree closely over the whole turn.

\subsection{Accuracy of the MLIP}

\Cref{tab:barriers} quantifies the comparison. Four observations summarize it.

\begin{table}[!htb]
  \centering
  \caption{Rotational barriers (\cref{eq:barrier}) from DFT and MACE-OFF23 and the agreement between the full profiles
  (\cref{eq:metrics}); $\delta_{\mathrm{rel}}=(\dE_{\mathrm{MACE}}-\dE_{\mathrm{DFT}})/\dE_{\mathrm{DFT}}$. MAE and RMSE are in eV;
  $r$ is the Pearson correlation coefficient. $\SI{1}{eV}=\SI{96.485}{kJ.mol^{-1}}$.}
  \label{tab:barriers}
  \small
  \setlength{\tabcolsep}{5pt}
  \begin{tabular}{@{}lcccccccccc@{}}
    \toprule
     & & \multicolumn{2}{c}{$\dE$ (eV)} & \multicolumn{2}{c}{$\dE$ (kJ\,mol$^{-1}$)} & & & & \\
    \cmidrule(lr){3-4}\cmidrule(lr){5-6}
    Mode & $\Delta\varphi_{\mathrm{step}}$ & DFT & MACE & DFT & MACE & $\delta_{\mathrm{rel}}$ (\%) & MAE & RMSE & $r$ \\
    \midrule
    $\alpha$ & $5^\circ$ & 2.948 & 2.856 & 284.5 & 275.6 & $-3.1$ & 0.132 & 0.257 & 0.948 \\
    $\ab$ & $5^\circ$ & 0.669 & 0.537 & 64.5 & 51.8 & $-19.7$ & 0.052 & 0.065 & 0.967 \\
    $\beta$ & $5^\circ$ & 0.807 & 0.657 & 77.9 & 63.4 & $-18.6$ & 0.063 & 0.075 & 0.965 \\
    $\gamma$ & $1^\circ$ & 0.178 & 0.162 & 17.1 & 15.6 & $-8.9$ & 0.007 & 0.010 & 0.984 \\
    $\delta$ & $1^\circ$ & 0.063 & 0.060 & 6.0 & 5.8 & $-4.2$ & 0.005 & 0.006 & 0.952 \\
    \bottomrule
  \end{tabular}
\end{table}

\begin{enumerate}[label=(\arabic*),leftmargin=2em,itemsep=2pt,topsep=3pt]
  \item \emph{Ordering and shape.} MACE reproduces the ordering of the barriers,
        \[\dE_\alpha\gg\dE_\beta\gtrsim\dE_{\ab}\gg\dE_\gamma>\dE_\delta,\]
        and the shape of every profile, with Pearson coefficients between 0.95 and 0.98 and global maxima displaced by at most
        $10^\circ$.
  \item \emph{Systematic underestimation.} All five barriers are lower in MACE than in DFT. The relative deviation is small for the
        $\alpha$ ($-3\%$), $\delta$ ($-4\%$) and $\gamma$ ($-9\%$) modes and largest for the acrylate rotations ($\approx-19\%$,
        i.e., \SIrange{0.13}{0.15}{eV}).
  \item \emph{Noise in the reference.} Part of the deviation for the acrylate modes comes from the reference itself. The DFT
        profiles for $\ab$ and $\beta$ are visibly rough: their root-mean-square deviation from a three-point running mean is
        \SI{0.032}{eV} and \SI{0.017}{eV}, against \SI{0.004}{eV} and \SI{0.007}{eV} for MACE. Neighbouring DFT points thus
        appear to end in slightly different local minima, and taking the maximum of a noisy profile biases $\dE_{\mathrm{DFT}}$
        upwards. A further source of systematic difference is the level of theory: MACE-OFF23 reproduces
        $\omega$B97M-D3(BJ)/def2-TZVPPD data~\cite{eastman2023spice,kovacs2025off}, which is not the level of the GPAW reference.
  \item \emph{The discontinuity of the $\alpha$ profile.} The MAE and RMSE of the $\alpha$ mode are dominated by the one-step offset
        of the abrupt drop: at $\Delta\varphi=235^\circ$ DFT is still at the maximum while MACE has already dropped, a difference of
        \SI{1.9}{eV} at a single point. Without the two points adjacent to the drop, the MAE and RMSE fall to \SI{0.105}{eV} and
        \SI{0.128}{eV}.
\end{enumerate}

Overall, a foundation model used without any system-specific training reproduces the qualitative and semi-quantitative features
of the rotational energy landscape of AESO. For the small rotors ($\gamma$, $\delta$) the agreement is quantitative; for the acrylate
rotations the barriers should be regarded as lower bounds with an uncertainty of about \SI{0.15}{eV}.

\subsection{Connection to dielectric relaxation}

For a thermally activated, non-cooperative reorientation, the relaxation time follows the Arrhenius law
\begin{equation}
  \tau = \tau_0\exp\!\left(\frac{E_{\mathrm{a}}}{k_{\mathrm{B}}T}\right),\qquad f_{\mathrm{p}}=\frac{1}{2\pi\tau},
  \label{eq:arrhenius}
\end{equation}
where $\tau_0$ is an attempt time of the order of an inverse molecular vibration frequency, which we take as
\SI{e-13}{s}, and $f_{\mathrm{p}}$ is the frequency of the corresponding loss peak~\cite{kremer2003}. Inverting \cref{eq:arrhenius} for the measurement
window of \SIrange{500}{5000}{Hz} at $T\approx\SI{295}{K}$ ($k_{\mathrm{B}}T=\SI{0.0254}{eV}$) gives
$E_{\mathrm{a}}=k_{\mathrm{B}}T\ln[1/(2\pi f_{\mathrm{p}}\tau_0)]\approx\SIrange{0.50}{0.56}{eV}$; changing $\tau_0$ by an order of
magnitude in either direction widens this range to \SIrange{0.44}{0.62}{eV}. Three consequences follow.

\begin{itemize}[leftmargin=1.6em,itemsep=2pt,topsep=3pt]
  \item The acrylate rotations ($\ab$ and $\beta$; \SIrange{0.54}{0.81}{eV} depending on mode and method) lie in or just above this
        window. They are the best candidates for the orientational defects that dominate the response at
        \SIrange{500}{5000}{Hz}. This agrees with the assignment of the $\ab$ and $\beta$ processes to intermediate frequencies and to a
        large contribution to dielectric loss (\cref{sec:od}).
  \item The methyl and hydroxyl rotations ($\gamma$, $\delta$; \SIrange{0.06}{0.18}{eV}) correspond to relaxation times of the order
        of \SIrange{e-12}{e-10}{s}, i.e., to loss maxima in the GHz--THz range. At the frequencies of our experiment these
        rotors are effectively instantaneous and contribute only to the high-frequency permittivity.
  \item The $\alpha$ profile does not describe a single activated jump. Its \SI{2.9}{eV} maximum is an upper bound produced by the
        constraint, and the real segmental relaxation is cooperative. In the curing network it is progressively frozen as the
        glass transition temperature rises above the measurement temperature.
\end{itemize}

These estimates are deliberately rough: the computed quantities are intramolecular rotational barriers of an isolated fragment,
whereas activation energies measured by DRS also include the contribution of the surrounding chains and free volume, and $\tau_0$ is
not known independently. Detailed DRS data for AESO:VDM are not yet available to us. Moreover, identical values could not be
expected, because the measured activation energies depend on the experimental conditions, under which statistically different
environments of a given OD can be realized. Nevertheless, the activation energies of the main ODs reported for structurally
related polymers---poly(methyl methacrylate) (PMMA) and poly(methyl acrylate) (PMA)---are close to the values calculated
here~\cite{sengwa2014,dedeus2004,suzuki2019,svajdlenkova2018}. We expect that a full analysis of the dielectric spectra of
AESO:VDM, combined with the structural changes during curing, will allow the theoretical results to be refined and extended.

\subsection{Limitations and outlook}

The present calculations treat a single AESO fragment. Neighbouring chains, cross-links and the free volume of the forming network,
all of which modify the barriers in the real material, are not included explicitly. The scans are sequential relaxed scans. They
are therefore not strictly periodic: the energies at $0^\circ$ and $360^\circ$ differ by up to \SI{0.15}{eV} for the $\ab$ mode,
and the abrupt drop of the $\alpha$ profile shows that the system can switch between basins along the path. Natural next steps are
(i)~nudged-elastic-band or free-energy calculations, which locate true saddle points and include entropic contributions,
(ii)~fine-tuning of the MACE model on reference data for AESO and its oligomers, which would remove the systematic offset in
\cref{tab:barriers}, and (iii)~molecular dynamics of cross-linked AESO:VDM models at finite temperature, from which relaxation
times---and eventually dielectric spectra---can be obtained directly and compared with experiment.

\section{Conclusions}\label{sec:conclusions}

We have modelled experimentally, and analysed, the photo-stimulated polymerization of AESO:VDM in an electric field, using an
intermittent illumination protocol that captures both fast curing and slow dielectric relaxation. The tools of complex-systems
theory, namely the sliding-window Shannon entropy and correlation integral, identify structural transformations of different
degrees of collectivity and separate them in time. Short light pulses at the beginning of curing trigger strongly correlated
relaxation, which fades as the network forms, leaving only localized orientational defects active.

The concept of orientational defects, which originates in the physics of ice and covalent crystals, provides a unified language
for these localized processes in polymers. Using MLIPs, specifically the MACE models~\cite{batatia2025mp,kovacs2025off}, we
computed relaxed rotational profiles for five orientational defects of AESO. Without any system-specific training, MACE-OFF23
reproduces the DFT profiles with correlation coefficients of 0.95--0.98 and the barrier hierarchy from \SI{2.9}{eV} ($\alpha$,
constrained) through \SIrange{0.54}{0.81}{eV} (acrylate rotations) to \SIrange{0.06}{0.18}{eV} (methyl and hydroxyl rotations). It
underestimates the barriers systematically by 3--20\%, most for the acrylate group. An Arrhenius estimate links the acrylate
rotations to the kHz window of the dielectric measurement and places the small rotors in the GHz--THz range.

AESO is a bio-based polymer derived from renewable resources and is attractive for sustainable materials development. The
combination of complexity analysis of in-situ measurements with MLIP-based modelling of the underlying defects, demonstrated here,
offers an efficient route to the energy barriers and relaxation profiles that govern its behaviour during photopolymerization.
We expect such insight to contribute to the rational design and optimization of AESO-based polymers for various applications.

\section*{Data availability}
The data supporting the findings of this study are available from the corresponding authors on reasonable request.

\section*{Acknowledgments}
This work was supported in part by the Ministry of Education and Science of Ukraine (projects Nos.~0122U000850, 0122U000874,
0122U001694, 0123U103572, 0124U003406, 0125U002005 and 0125U002033), National Research Foundation of Ukraine (project
No.~2020.02/0100), Slovak Grant Agency VEGA (projects Nos.~2/0134/21, 2/0166/22, and 2/0131/25), and Slovak Research and Development
Agency (project No.~APVV-21-0335). T.K. also acknowledges the SAIA (Slovak Academic Information Agency) for scholarship in the
Institute of Physics of Slovak Academy of Sciences in the framework of the National Scholarship Programme of the Slovak Republic.
This work has also received funding through the MSCA4Ukraine project (grant No.~1128327), which is funded by the European Union,
and the EURIZON project (grant EU-3022), which is funded by the European Union (EURIZON H2020 project) under grant agreement
No.~871072. The authors would also like to thank the Armed Forces of Ukraine for providing security to perform this work. This work
has become possible only because of the resilience and courage of the Ukrainian Army.



\end{document}